\documentclass[final,authoryear,5p]{elsarticle}

\usepackage{graphicx}
\usepackage{subfigure}
\usepackage{verbatimbox}
\usepackage{amsmath}	
\usepackage{amssymb}	

\usepackage[
a4paper=true,%
breaklinks=true,%
colorlinks=true,%
pdfauthor={First Author et al.},%
pdftitle={Template for manuscripts in Advances in Space Research}%
]{hyperref}

\journal{Advances in Space Research}

\newcommand*{\restrictlinewidthbox}[1]{%
  \begingroup
    \sbox0{#1}%
    \ifdim\wd0>\linewidth
      \resizebox{\linewidth}{!}{\copy0}%
    \else
      \copy0 %
    \fi
  \endgroup
}
\newcommand{\be}{\begin{equation}}
\newcommand{\ee}{\end{equation}}

\begin{document}

\begin{frontmatter}









\title{A fast algorithm for the detection of faint orbital debris tracks in optical images}

\address[address1]{Department of Physics and Astronomy, The University of British Columbia, 6224 Agricultural Road, Vancouver, BC, V6T1Z1, Canada}
\author[address1,address2,address3]{P. Hickson\corref{cor}} 
\address[address2]{Space science, Technologies and Astrophysics Research (STAR) Institute,Universit\'e de Li\`ege, Institut d$'$Astrophysique et de G\'eophysique, All\'ee du 6 Ao\^ut 19c, 4000 Li\`ege, Belgium}
\address[address3]{National Astronomical Observatories, Chinese Academy of Sciences,  20A Datun Road, Chaoyang District, Beijing, China}
\cortext[cor]{Corresponding author}
\ead{hickson@physics.ubc.ca}

\begin{abstract}
Moving objects leave extended tracks in optical images acquired with a telescope that is tracking stars or other targets. By searching images for these tracks, one can obtain statistics on populations of space debris in Earth orbit. The algorithm described here combines matched filtering with a Fourier implementation of the discrete Radon transform and can detect long linear tracks with high sensitivity and speed. Monte-Carlo simulations show that such tracks, in a background of Poisson random noise, can be reliably detected even if they are invisible to the eye. On a 2.2 GHz computer the algorithm can process a $4096\times 4096$-pixel image in less than a minute.
\end{abstract}


\end{frontmatter}

\parindent=0.5 cm

\let\thefootnote\relax\footnote{\copyright 2018. This manuscript version is made available under the CC-BY-NC-ND 4.0 license http://creativecommons.org/licenses/by-nc-nd/4.0/ }

\section{Introduction}

The detection of linear tracks in a two dimensional image is a common problem in image processing. One important application is for the detection of orbital debris. While known objects can be tracked with a telescope, increasing the signal-to-noise ratio, unknown objects cannot. An appropriate observing strategy is to track at the sidereal rate, thereby minimizing image contamination by stars, and search for tracks in the image produced by objects moving across the field of view during the exposure. Automated algorithms can process large data sets and can reach detection limits fainter than can human observers. 

Many groups have developed algorithms to find streaked images, for the detection of moving celestial objects \citep{sara2017,waszczak2017} as well as for satellite or debris detection \citep{zimmer2013,ciurte2014,vananti2015,virtanen2017,vallduriola2018}. A variety of techniques have been employed. In segmentation-based methods \citep{liu1992,virtanen2017},  pixels having intensities above a threshold are analyzed. This is well-suited to short, relatively-bright streaks. Stacking methods are useful when an object is observed in multiple images \citep{yanagisawa2012}. Other methods employ the Radon or Hough transform \citep{zimmer2013,ciurte2014} or matched filtering \citep{gural2005,schildknecht2015,sara2017}. 

The classical Radon transform \citep{radon1917,radon1986}, and its close relative the Hough transform \citep{hough1959,duda1972}, have long been employed to identify linear features in images. These transforms map lines to points in a two-dimensional ``Hough space'', whose axes correspond to position and angle. The angle is that between the normal to the line and the reference axis, and the position is the perpendicular distance from the line to the origin, usually taken to be the centre of the image. Thus the problem of detecting long streaks becomes a simpler problem of detecting local maxima in Hough space. A great advantage of this approach is that fast techniques  have been developed for the computation of the Radon transform \citep{beylkin1987,gotz1995,press2006}. These are fast in the sense that for an $N \times N$ image, processing time grows roughly in proportion to $N^2 \ln N$, rather than $N^3$ for the classical Radon transform. For a typical astronomical CCD image, $N \sim 10^3 - 10^4$, so the fast Radon transform requires typically two to three orders of magnitude fewer computations.

This paper describes a method that combines matched filtering with a fast discrete Radon transform in order to achieve high sensitivity and speed. It is best suited for the detection of long faint streaks in a single image, as would be produced by fast-moving objects. The algorithm was tested by Monte Carlo simulation of random linear tracks, having a Gaussian cross-section of specified FWHM, superimposed on a constant background, to which was then added random Poisson noise. It was found to be capable of reliably detecting faint tracks that were invisible to the eye. 

The sensitivity of the method comes from the use of matched filtering, which provides the highest possible signal-to-noise ratio of any linear detection technique, allowing the faintest possible detection limit. Direct application of matched filtering, by integrating along all possible directions and positions in the image, would be equally sensitive but very slow. The Radon transform decreases the number of dimensions of the search, providing a large increase in speed.

We begin by briefly reviewing the concepts of optimal detection and the Radon transform. The algorithm is then described and results of the simulations are presented. Our python source code implementing the Radon transform is reproduced in Appendix A.

\section{Method}

\subsection{Optimal detection}

It has been known since 1953 that the optimal linear technique for the detection and measurement of a signal in the presence of uniform uncorrelated stochastic noise is that of matched filtering \citep{woodward1953,turin1960}. It is optimal in the sense that no other linear filter can give a higher signal-to-noise ratio.

In order to apply this method to the problem of detecting and measuring faint tracks in noisy images, let us first suppose that we know a priori the angle that the track makes with one of the axes of the image. Then, we can integrate along this direction, summing the pixel values along lines parallel to the track, in order to produce a one-dimensional mean profile of the cross-section of the track. Effectively, this amounts to \emph{projecting} the image onto a line that is orthogonal (transverse) to the track. In order to measure the position of the track, and the total flux that it contains, we can search this one-dimensional projection for a local maximum.

If we further know the shape and width of the track, in the transverse direction, we can employ matched filtering. This will generally be true as tracks left by orbital debris are typically unresolved, having a transverse intensity profile that is well-approximated by the one-dimensional projected profile of the point-spread function (PSF), found from images of stars in the field. Denote the summed intensity along the projection line by $p(r)$, where $r$ is a coordinate in the direction transverse to the track. Assume that any constant intensity $I_0$, such as the sky background, has first been subtracted. Let $f(r)$ be the expected profile, i.e. the projected PSF, normalized so that its one-dimensional integral is unity,
\be
  \int f(r) dr = 1. \label{eq:norm1}
\ee
For simplicity we show here the results for continuous functions and the integral extends over the entire domain of the function. The extension to discrete values can be made by replacing integrals by summations. 

If the track has the expected profile, and is centred at $r = r_0$, we may write
\be
  p(r) = F f(r-r_0) + n(r), \label{eq:prof}
\ee
where $F$ is the total flux in the track (the integral along the projection line of the summed intensity in the track) and $n(r)$ is a random variable having zero mean and variance $\sigma^2_n(r)$, representing the noise. Normally this will be the single-pixel noise variance $\sigma_p^2$ multiplied by the number of pixels that were summed in each line parallel to the track. 

To detect the track with optimal sensitivity, we cross-correlate the summed intensity profile with a function $h(r)$ that is proportional to the expected signal divided by the noise variance \citep{king1983},
\be
  h(r) = \alpha \frac{f(r)}{\sigma^2_n(r)}. \label{eq:h}
\ee
The constant $\alpha$ is chosen to make the integral of $h(r)$ unity,
\be
  \int h(r) dr = 1, \label{eq:norm2}
\ee
The cross correlation will have a maximum at the location of the track, where it takes the value
\begin{align}
  g(r_0) & =  F \int h(r) f(r) dx + \int h(r) n(r) dr. \label{eq:g}
\end{align}
This is a fluctuating quantity having an ensemble average 
\be
  \overline{g(r_0)} =  F \int f(r) h(r) dr.
\ee
The second term has disappeared by virtue of $n(r)$ having zero mean. The best estimate of the true flux $F$ is therefore
\be
  \hat{F} = \frac{g(r_0)}{Q}, \label{eq:fhat}
\ee
where
\be
  Q = \int f(r) h(r) dr. \label{eq:Q}
\ee

The variance of this estimate is
\begin{align}
  \text{Var } \hat{F} & = \frac{1}{Q^2} \text{Var } g(r_0), \nonumber \\
  & = \frac{1}{Q^2} \int h^2(r) \sigma^2_n(r) dr, \nonumber \\
  & = \frac{\alpha}{Q},
\end{align}
so the signal-to-noise ratio is
\be
  s = F \sqrt{\frac{Q}{\alpha}}. \label{eq:s1}
\ee

If the noise variance can be assumed to be constant, independent of $r$, then Eqns. (\ref{eq:norm1}), (\ref{eq:h}) and (\ref{eq:norm2})
require that $\alpha = \sigma_n^2$, and therefore $h(r) = f(r)$. The matched filter is proportional to the expected signal. In that case,
\be
  Q = \int f^2(r) dr,
\ee
and the signal-to-noise ratio (SNR) becomes
\be
  s = \frac{F}{\sigma_n}\sqrt{Q} =  \frac{\overline{g(r_0)}}{\sigma_n\sqrt{Q}}. \label{eq:s2}
\ee
This shows the importance of a sharp PSF (small FWHM), which increases $Q$, improving the SNR.

Although the noise in astronomical images arises from a number of different sources \citep{newberry1991,howell2003}, it is often dominated by the Poisson statistics of the detected photons. In that case, the noise variance will be
\be
  \sigma^2_n(r) = F_0 + F f(r),
\ee
where $F_0$ represents the projected intensity of the background light, before sky subtraction. Even though the mean background has been subtracted, its noise remains.

For the Poisson case, the optimal filter, Eqn. (\ref{eq:h}), becomes
\be
  h(r) = \frac{\alpha}{F_0} \frac{f(r)}{1+\beta f(r)},
\ee
where
\be
  \alpha = F_0 \left[ \int \frac{f(r) dr}{1+\beta f(r)} \right]^{-1}.
\ee
Here $\beta = F/F_0$ is a measure of the relative brightness of the track compared to the background. The constant $Q$ is now
\be
  Q = \frac{\alpha}{F_0} \int \frac{f^2(r) dr}{1+\beta f(r)},
\ee
so the SNR, Eqn. (\ref{eq:s1}), becomes
\be
  s = \frac{F}{\sigma_0}\left[ \int \frac{f^2(r) dr}{1+\beta f(r)}\right]^{1/2}. \label{eq:s3}
\ee
where $\sigma_0^2 = F_0$ is the variance of the background.

For the Poisson case we see that the optimal filter depends on the flux of the track, which is generally not known in advance. But, the problem considered in this paper is the efficient detection of \emph{faint} tracks in a noisy image. In that case, $\beta \lesssim 1$, and the Poisson equations approach those of the constant-variance case, as  expected. For bright streaks, it does not matter if the matched filter that is used is somewhat less than optimal. They will be detected in any case. This is primary justification for employing the constant-variance matched filter when searching for faint tracks. 

\subsection{The fast discrete Radon transform}

Of course if one knew in advance the orientation of the track, the detection problem would be relatively simple. But in general the orientation is not known. Thus, the algorithm must search all possible orientations. Directly computing projections for $\sim N$ orientations is time consuming. However, the speed of the process can be increased greatly by the use of the Fast Radon Transform. 

The algorithm that we employ to compute the Radon transform of the image is based on the Fourier Slice Theorem \citep{bracewell1956}. This theorem, which is easily proved, states that the values on a slice through the origin of the two-dimensional Fourier transform of the image is equal to the one-dimensional Fourier transform of the projection of the image onto a line parallel to the slice. This allows one to employ fast Fourier transforms to compute the Radon transform, by taking the inverse Fourier transform of each slice, for a complete set of angles.

One way to compute the values on the slice would be to use two-dimensional interpolation in the transformed image to estimate the values at integer distances (in units of pixels) along the slice. However, a simpler and faster method is employed here. If the angle $\theta$ between the slice and the $x$ axis is in the range $|\theta| <= 45^\circ$ (for square images), a value on the slice is determined for every $x$ pixel by taking the value of the image at $(x,y)$, where $y = x\tan\theta$. Sinc interpolation is used to estimate the value of the transformed image at fractional values of $y$. In this way, the problem of interpolation in two dimensions is reduced to one-dimensional interpolation. The resulting intensities along the slice have a spacing of $x/\cos\theta$. To compensate, according to the Fourier scaling theorem the intervals of $r$ after taking the inverse Fourier transform, are multiplied by a factor of $\cos\theta$ and the intensity is multiplied by a factor of $\sec\theta$. If $|\theta| > 45^\circ$, the approach is similar but with $x$ and $y$ interchanged and $\tan\theta$ replaced by $\cot\theta$ and $\cos\theta$ replaced by $\sin\theta$.

This differs from the standard Radon transform in that the position coordinate no longer measures perpendicular distance from the track to the origin, but distance along the $x$ or $y$ axis, depending on the value of $\theta$. The matched filter is easily adapted to this by scaling the FWHM of the PSF by a factor of $\sec\theta$ (or $\csc\theta$ if $|\theta| > 45^\circ$) in order to account for the oblique cut through the track.

The interpolation scheme that is used to compute values on the slice is important. Simple linear interpolation, or even polynomial interpolation, produces artifacts in the Radon transform, which degrade the photometric accuracy. The correct approach is to use sinc interpolation. Here there is a tradeoff between the order of the interpolation (the number of pixels that are included in the summation) and the speed of the technique. A full $N-$point sinc filter completely eliminates the artifacts, but significantly increases processing time. On the other hand, linear interpolation, which is very fast, results in systematic photometric errors that can be as great as 15\%. Employing a sinc interpolating filter encompassing 7 pixels reduces photometric errors to less than 5\%. Increasing this to $\sim 50$ pixels reduces photometric errors to less than 1\%, but increases execution time by about a factor of 4. Even so, this is still more than an order of magnitude faster than the corresponding two-dimensional interpolation.

\subsection{The algorithm}

Our method involves the following steps:

\begin{itemize}
\item If the image is not square, divide it into overlapping square sub-images. Then for each sub-image:
\item Mask stars and image defects.
\item Subtract the median background.
\item Compute the Radon transform.
\item Determine the RMS noise $\sigma_n$ in the Radon image and the threshold value of $g$ corresponding to the desired SNR limit (from Eqn \ref{eq:s2}).
\item Find the highest value in the Radon image and record the corresponding position, angle, flux and SNR.
\item Mask the region around the highest value bounded by specified tolerances in position and angle.
\item Repeat this, finding the highest value in the masked Radon image, and continue until the highest value falls below the threshold.
\item Combine the detections for all sub-images and reject duplicate detections.
\end{itemize}

The procedure requires some judgement about what constitutes a duplicate detection. This is best found from experience, but typically, two detections that have positions within a few FWHM of each other and angles within two or three degrees are considered equivalent and the detection with the highest flux is selected. In some cases, ``ghost'' detections may occur, which have the same angle but differ in position by the number of pixels along an axis of the image. This results from the periodicity of the fast Fourier transform.

\section{Simulations and results}

\begin{figure}[htb!]
\centering
\begin{subfigure}{}
\centering
\includegraphics[width=0.8\columnwidth]{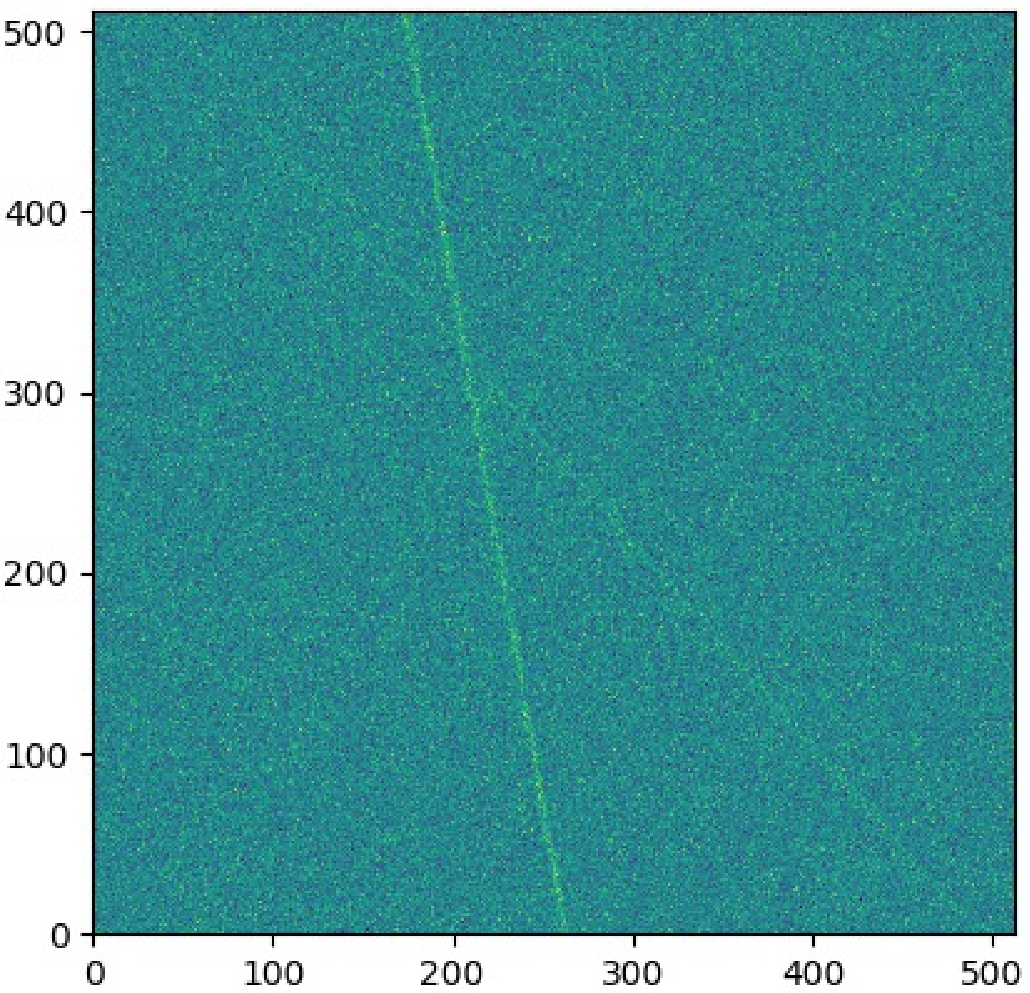}
\end{subfigure} \\
\begin{subfigure}{}
\centering
\includegraphics[width=\columnwidth]{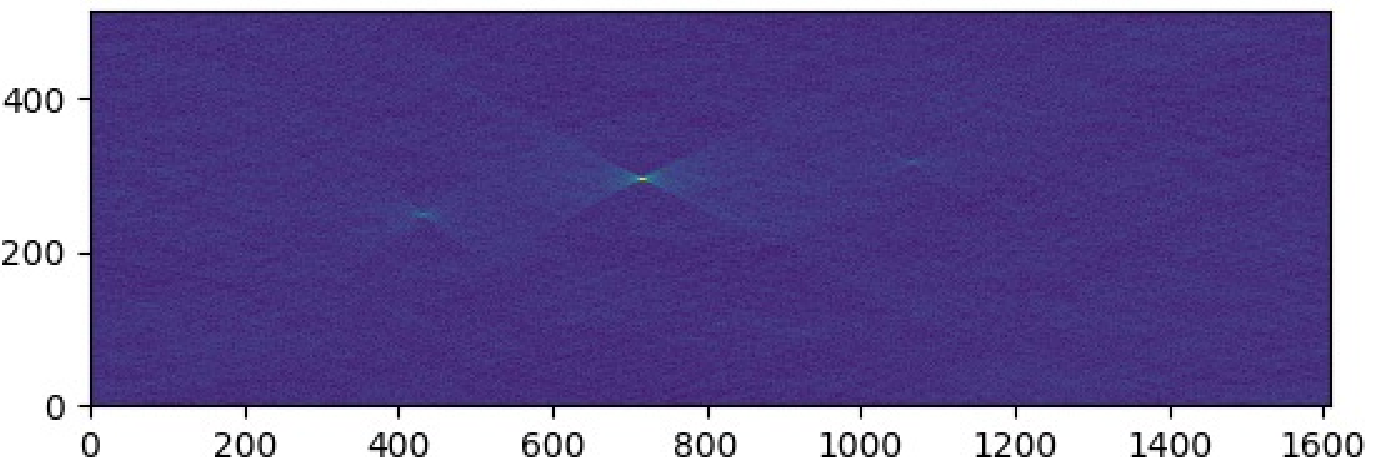}
\end{subfigure}
\vspace{-6pt}
\caption{Simulated image containing three orbital debris tracks (upper) and its Radon transform (lower). The horizontal scale on the lower image is pixel number, but it represents a range of angles from $-90^\circ$ to $90^\circ$. The vertical scale is the $x$ position of the midpoint of the streak. The tracks have a FWHM of 3.0 pixels and signal-to-noise ratios of 100, 50 and 25. Tracks having a signal-to-noise ratio of $\sim 20$ or less are generally invisible to the eye in a 1K$\times$1K or larger image, but are nevertheless detected by the algorithm.}
\label{fig:sim}
\end{figure}

The algorithm was coded in Python 3 and uses the Numpy and Scipy libraries. In order to test it, an image was created and filled with random values from a standard normal distribution ($\sigma_p = 1$). Then, a streak was constructed having a random orientation and a transverse profile given by a Gaussian having standard deviation $\sigma = w/\sqrt{8\ln 2}$, where $w$ is the desired FWHM in pixels. The profile was normalized so that its integral, multiplied by the number of pixels along the length of the track, equals $F$, the flux required to achieve the desired signal-to-noise ratio according to Eqn. (\ref{eq:s2}). An example of a simulated image containing several tracks, and the corresponding Radon transform, is shown in \autoref{fig:sim}. 

A total of 2700 simulations were run using a range of SNR and FWHM. The results are summarized in \autoref{tab:monte_carlo}. Here FWHM is measured in pixels, and the success rate is the fraction of runs for which the strongest detection found by the algorithm matched the simulated track. The last column lists the magnitude error, defined by $-2.5\log_{10} (F_\text{measured}/F_\text{true})$.

The SNR values listed in \autoref{tab:monte_carlo} are computed using Eqn (\ref{eq:s2}), where $F$ is the modelled total flux of the track and $\sigma_n = \sqrt{F_0}$ is the background Poisson noise variance.

\begin{table}[htb!]
\centering
\caption{Monte-Carlo simulation summary}
\label{tab:monte_carlo}
\vspace{6pt}
\resizebox{\columnwidth}{!}{
\begin{tabular}{lrrrrr} 
\hline \\[-9pt]
SNR & Size & No of trials & FWHM & Success rate & Mag error\\
\hline \\[-9pt]

2.0 & 1024 & 1000 & 2.0 & 0.001 & 1.15 \\
 &  &  & 3.0 & 0.002 & -1.02 \\
 &  &  & 4.0 & 0.001 & 1.01 \\
 
3.0 & 1024 & 1000 & 2.0 & 0.008 & 0.76 \\
 &  &  & 3.0 & 0.011 & 0.66 \\
 &  &  & 4.0 & 0.027 & 0.66 \\
 
4.0 & 1024 & 1000 & 2.0 & 0.075 & 0.39 \\
 &  &  & 3.0 & 0.094 & 0.42 \\
 &  &  & 4.0 & 0.125 & 0.37 \\

5.0 & 1024 & 1000 & 2.0 & 0.255 & 0.20 \\
 &  &  & 3.0 & 0.328 & 0.22 \\
 &  &  & 4.0 & 0.370 & 0.22 \\
 
6.0 & 1024 & 1000 & 2.0 & 0.620 & 0.13 \\
 &  &  & 3.0 & 0.642 & 0.14 \\
 &  &  & 4.0 & 0.687 & 0.15 \\
 
7.0 & 1024 & 1000 & 2.0 & 0.859 & 0.14 \\
 &  &  & 3.0 & 0.879 & 0.13 \\
 &  &  & 4.0 & 0.914 & 0.14 \\
 
8.0 & 1024 & 1000 & 2.0 & 0.964 & 0.15 \\
 &  &  & 3.0 & 0.971 & 0.14 \\
 &  &  & 4.0 & 0.979 & 0.13 \\
 
9.0 & 1024 & 1000 & 2.0 & 0.997 & 0.15 \\
 &  &  & 3.0 & 0.999 & 0.14 \\
 &  &  & 4.0 & 0.997 & 0.13 \\
 
10.0 & 1024 & 1000 & 2.0 & 0.999 & 0.14 \\
 &  &  & 3.0 & 1.000 & 0.13 \\
 &  &  & 4.0 & 1.000 & 0.12 \\
 
\hline
\end{tabular}
}
\end{table}

\begin{figure}[htb!]
\centering
\hbox{\hspace{-0.7cm}\includegraphics[width=0.57\textwidth]{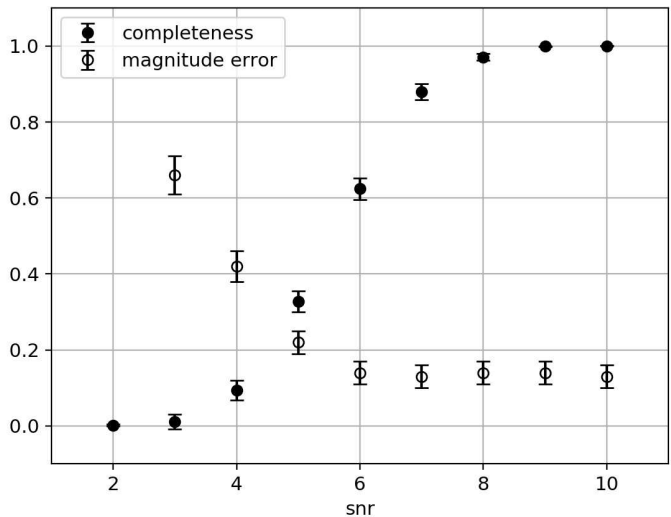}}
\vspace{-6pt}
\caption{Completeness and photometric error vs. signal-to-noise ratio, for tracks having a FWHM of 3.0 pixels.}
\label{mc}
\end{figure}

\section{Effect of track curvature}

The algorithm is designed to find linear tracks. Tracks that are slightly curved will still be detected, but with lower sensitivity. Tracks produced by orbital debris are not perfectly straight, although the deviation from linearity over the field of view of a typical astronomical camera is generally quite small. 

An exact analysis of the curvature of debris tracks is beyond the scope of this paper, but a simple approximate treatment will suffice to provide an estimate magnitude of the effect. Long tracks are made by fast-moving debris in low or middle Earth orbit, which cross a typical imager in a few minutes or less. For such objects, there is little error in ignoring the motion of the observer due to the rotation of the Earth. Also, for simplicity, we shall assume that the orbit is circular.

\begin{figure}[htb!]
\centering
\includegraphics[width=0.49\textwidth]{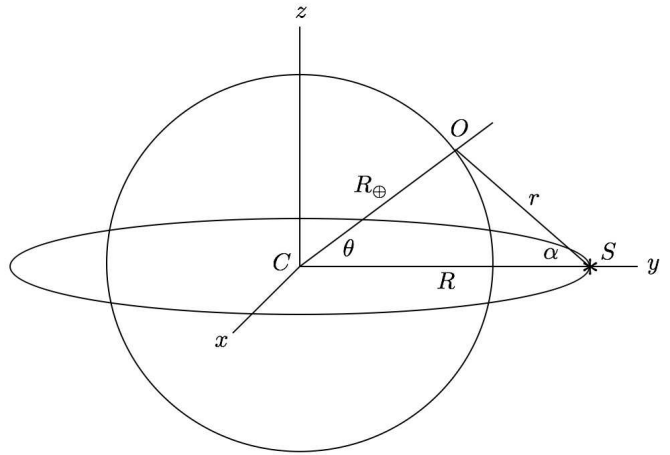}
\vspace{-12pt}
\caption{Geometry for estimating track curvature. A satellite at point $S$ in a circular orbit is observed from point $O$ on the surface of the Earth. The track appears elliptical when projected perpendicular to the line of sight.}
\label{fig:orbit}
\end{figure}

The relevant geometry is shown in Figure \ref{fig:orbit}. We choose a barycentric Cartesian coordinate system in which the object orbits in the $x-y$ plane. Consider an observer $O$ viewing the orbiting object when it is highest in the sky, which is the point where it crosses the $x-z$ axis. The observer sees the orbit projected on the sky, where it appears to be elliptical. The apparent curvature of the track (the reciprocal of the angular radius of curvature in radians) is
\be
  \kappa = \frac{r}{R}\sin\alpha = \frac{R_\oplus}{R}\sin\theta, \label{eq:kappa}
\ee
where $R$ is the orbital radius, $R_\oplus$ is the radius of the Earth, $r$ is the distance from the observer to the object, $\alpha$ is the angle between the line of sight and the orbital plane, and $\theta$ is the angle between the line connecting the observer to centre and the orbital plane. The second equality follows by application of the sine rule for plane triangles.

From this we see that for a given angle $\theta$, the curvature is maximized by making $R$ as small as possible. The smallest possible value is $R = R_\oplus/\cos\theta$, which places the object on the observer's horizon. With this choice of $R$, Equation (\ref{eq:kappa}) becomes
\be
  \kappa = \frac{1}{2}\sin 2\theta, \label{eq:kappa2}
\ee
which has a maximum value $\kappa_\text{max} = 1/2$ which occurs when $\theta = \pi/4$.

For a track segment of length $\beta$ radians, the maximum angular deviation from the best-fit straight line is
\be
  \epsilon = \frac{1}{16}\kappa \beta^2.
\ee
In order for there to be no significant loss of sensitivity, this deviation should be smaller than the half-width of the PSF. For example, if the PSF has a FWHM of 1 arcsec, the maximum track length is 0.5 degrees. Longer tracks will spread beyond the extend of the matched filter, lowering the detection sensitivity. A lower limit on the sensitivity can be obtained by assuming that the outer regions of the track, beyond this maximum length, contribute nothing to the detection. In that case, the signal-to-noise ratio for the detection of a maximally-curved 1-degree track would be reduced by a factor of two. In practice, the loss would be smaller than this. Also, this is the worst-case curvature. Most tracks should have much less deviation from linearity.

This analysis suggests that track curvature should not have a significant impact for imagers having a field of view less than one degree. However, the deviation increases quadratically with track length, so it is clear that curvature could be an issue for wide-field cameras having a larger field of view.

A second important source of nonlinearity is distortion within the telescope and camera optics. This distortion needs to be corrected to a fairly tight tolerance, on the order of 0.1\% or less, in order to prevent significant distortion of the tracks.

\section{Discussion}

It can be seen from \autoref{tab:monte_carlo} and \autoref{fig:sim} that the completeness depends primarily on the total signal-to-noise ratio. For a given SNR, There is no significant variation with the width of the track or with the intensity or flux of the track. The 50\% completeness limit corresponds to $SNR \sim 5.5$, and essentially all objects with $SNR \ge 8$ are detected. Larger images have more distinguishable positions and angles, so more opportunities for noise events to rise above the detection threshold. However this is compensated by longer tracks, which give a higher SNR for true events. In our simulations, tracks having a total $SNR \sim 20$ or less are generally invisible to the eye. Yet they are readily detected by the algorithm described here. 

The photometric errors found in these simulations are consistent with the expected Poisson noise. As a check for systematic errors, several simulations were run with very bright tracks, for which the Poisson noise was negligible. These had photometric errors that were less than 0.01 magnitudes (approximately 1\%) when 51-point interpolaton was used and 0.05 magnitudes for 7-point interpolation.

No attempt was made to simulate stars, which need to be masked before running this algorithm on astronomical images. Masking of stars can be done automatically, and the effect on streaks is generally quite small unless the field is very crowded \citep{zimmer2013}.

The method is most sensitive for the detection of tracks that completely cross the image. Tracks that end within the image can also be detected but with lower efficiency. The signal-to-noise ratio for such objects is proportional to the track length. Tracks that cross the image are suboptimal for the estimation of orbital and photometric parameters because the angular speed and intrinsic luminosity of the object cannot be determined unless both endpoints are contained within the image. Nevertheless, the relative brightness of the track, its orientation, and the time of passage all provide useful information.

The algorithm was implemented and tested on a computer having a 2.2 GHz 64-bit processor. Execution time was $\sim 3$ s for a 1K$\times$1K image, $\sim 10$ s for a 2K$\times$2K image and $\sim 30$ s for a 4K$\times$4K image. This speed could be increased by parallelizing the code in order to take advantage of multiple cores, however, it is already fast enough for many applications. For example, it could be useful for surveys such as that planned for the  International Liquid Mirror Telescope, which will regularly scan the sky at $29.36^\circ$ N latitude, acquiring a 16-Mpixel image every 102 seconds \citep{surdej2006}.  Such images can be scanned for streaks in near-real time in order to acquire statistics on orbital debris populations.

\section*{Acknowledgements}

I am grateful to Prof. J. Surdej for many discussions and a careful reading of the manuscript, and to the University of Li\`ege for hospitality during a sabbatical visit. This work was supported by grants from the Natural Sciences and Engineering Research Council of Canada and the Fonds de la Recherche Scientifique (FNRS) of Belgium, R.FNRS.4164-J-F-G. The hospitality of NAOC and support from the Chinese Academy of Sciences, via the CAS President’s International Fellowship Initiative, 2017VMA0013, is also gratefully acknowledged.

\section*{References}

\bibliography{references}

\begin{appendix}

\section{Source code for the fast Radon transform employed here.}

\setlength{\parindent}{0pt}
\begin{verbbox}
#------------------------------------------------------------------------------
# Fast 2-d Radon transform.
#
# Revision
#   2018-04-20
#
# Copyright (c) 2018 by P. Hickson. This code may be freely copied and modified
# for non-commercial applications only. This copyright notice must be retained
# in all copies and derivatives.
#
def radon(img,theta=np.arange(-90,90,1),order=7):
    """
    The function computes the 2d Radon transform of a square image. It uses
    the Fourier slice theorem. The 1-d inverse Fourier transform of a slice
    through the 2-d Fourier transform of the image is equal to the projection
    (sum) of the image along the slice axis. In this implementation, r is the
    lesser of the distance along the x and y axes, from the centre of the
    image to the streak. It is not the perpendicular distance as with the
    standard Radon transform. This choice was made to increase speed and
    reduce memory requirements.
    """

    # Define the slice function.
    def im_slice(im,th,order):
        """
        Return a slice through the origin along the specified direction. th is
        the angle between the slice and the x axis, i.e. th = 0 is a horizontal
        slice y = 0). The origin is located at (nx//2,ny//2). The range of th
        is -90 to +90 degrees.

        The parameter order is the order of the sinc interpoltion. Increase it
        for more accurate photometry, decrease it for faster speed.
        """
        ni = int(order)
        ni2 = ni//2
        ny,nx = im.shape
        x0 = nx//2
        y0 = ny//2
        tr = radians(th)
        ct = cos(tr)
        st = sin(tr)
        sl = np.zeros(nx,dtype=complex)
        x = np.arange(nx)
        y = np.arange(ny)
        if th < -45 or th > 45:
            xy = x0+(y-y0)*ct/st
            xmin = (xy-ni2).astype(int)
            for i in range(ni):
                xv = np.clip(xmin+i,0,nx-1)
                sl += im[y,xv]*np.sinc(xy-xv)
        else:
            yx = y0+(x-x0)*st/ct
            ymin = (yx-ni2).astype(int)
            for i in range(ni):
                yv = np.clip(ymin+i,0,ny-1)
                sl += im[yv,x]*np.sinc(yx-yv)
        return sl
        
    # Find the size of the image.
    ny,nx = img.shape
    if nx != ny:
        print('image must be square')
        exit(0)

    # Create an image to hold the output
    nt = len(theta)
    rt = np.zeros([nx,nt])

    # Take the 2-d FFT of the image.
    fimg = np.fft.fftshift(np.fft.fft2(np.fft.fftshift(img)))

    # For each angle, extract the slice, then take the 1-d inverse FFT.
    for i in range(nt):
        tr = radians(theta[i])
        sl = im_slice(fimg,theta[i],order)
        rt[:,i] = np.real(np.fft.ifftshift(np.fft.ifft(np.fft.ifftshift(sl))))
        if theta[i] < -45:
            rt[:,i] = rt[::-1,i]

    return rt
\end{verbbox}
\restrictlinewidthbox{\theverbbox}

\end{appendix}

\end{document}